\documentclass[a4paper]{article}

\usepackage{INTERSPEECH2022}

\usepackage{booktabs}
\usepackage{multirow}
\usepackage{array}
\usepackage{float}

\usepackage{tikz}
\usepackage{color}
\usepackage{kotex}
\usepackage{epsfig}
\usepackage{graphics}
\usepackage{subcaption}

\DeclareMathOperator*{\argmin}{arg\,min}
\DeclareMathOperator*{\softmax}{softmax}
\DeclareMathOperator*{\cossim}{sim}

\newcommand\sh[1]{\textcolor{brown} {#1}} 
\newcommand{\bg}[1]{\textcolor{blue} {#1}} 
\newcommand{\is}[1]{\textcolor{orange}{#1}} 

\renewcommand\sh[1]{\textcolor{black}{#1}} 
\renewcommand{\bg}[1]{\textcolor{black} {#1}} 
\renewcommand{\is}[1]{\textcolor{black}{#1}} 

\title{Personalized Keyword Spotting through Multi-task Learning}
\name{Seunghan Yang$^1$, Byeonggeun Kim$^1$, Inseop Chung$^{1,2,*}$, Simyung Chang$^1$}
\address{
  $^1$Qualcomm AI Research${}^{\dagger}$, Qualcomm Korea YH, Seoul, Republic of Korea  \thanks{  ${}^{\dagger}$ Qualcomm AI Research is an initiative of Qualcomm Technologies, Inc.${}^{*}$Author completed the research in part during an internship at Qualcomm Technologies, Inc.}\\
  $^2$Seoul National University, Seoul, Republic of Korea}
\email{\{seunghan, kbungkun, ichung, simychan\}@qti.qualcomm.com}

\begin{document}

\maketitle
\begin{abstract}
Keyword spotting (KWS) plays an essential role in enabling speech-based user interaction on smart devices, and conventional KWS (C-KWS) approaches have concentrated on detecting user-agnostic pre-defined keywords.
However, in practice, most user interactions come from target users enrolled in the device which motivates to construct personalized keyword spotting.
We design two personalized KWS tasks; (1) Target user Biased KWS (TB-KWS) and (2) Target user Only KWS (TO-KWS).  
To solve the tasks, we propose personalized keyword spotting through multi-task learning (PK-MTL) that consists of multi-task learning and task-adaptation.
First, we introduce applying multi-task learning on keyword spotting and speaker verification to leverage user information to the keyword spotting system.
Next, we design task-specific scoring functions to adapt to the personalized KWS tasks thoroughly.
We evaluate our framework on conventional and personalized scenarios, and the results show that PK-MTL can dramatically reduce the false alarm rate, especially in various practical scenarios.

\end{abstract}

\section{Introduction}

An always-on lightweight keyword spotting system has been exploited to wake up smart audio devices. When the system detects the keyword, the following audio stream can be uploaded to speech recognition systems~\cite{gruenstein2017cascade, res15, tcresnet, mittermaier2020small, zhang2020deep, zhang2021autokws}. This process can reduce power consumption while maintaining a high recall rate and low false alarm rate.
Conventional keyword spotting (C-KWS)~\cite{res15, bcresnet} aims to detect a small set of pre-defined speech signals such as wake-up words, {\it e.g.}, ``Alexa" and ``OK Google," and has been applied to the always-on keyword spotting system.

However, this system is not personalized, {\it i.e.}, it only focuses on pre-defined keywords and does not consider the users.
\sh{To adapt the system to the users, query-by-example keyword spotting~\cite{QBKWS_2, kim2019query, QBKWS_1} has been proposed to allow the users to enroll their own keywords, but these approaches only adapt to new keywords, not explicitly considering the user identity.}
In practice, most user interactions come from the \bg{\it{target}} users, and hence the model requires to be biased on the target users.
\sh{Moreover, since the aforementioned systems ignore user information, they cannot prevent detecting general negatives of background sounds containing target keywords or other keywords having similar pronunciation to the target, {\it e.g.}, streaming audios from TV, online meetings, and conversations.}
It could lead to undesirable power consumption by unnecessarily activating the recognition systems.
To address these problems, we introduce more practical, personalized KWS tasks, \sh{considering the target users enrolled in the device.}

\begin{figure}[t]
\centering
  \begin{subfigure}{1.0\linewidth}
  \centering
  \epsfig{figure=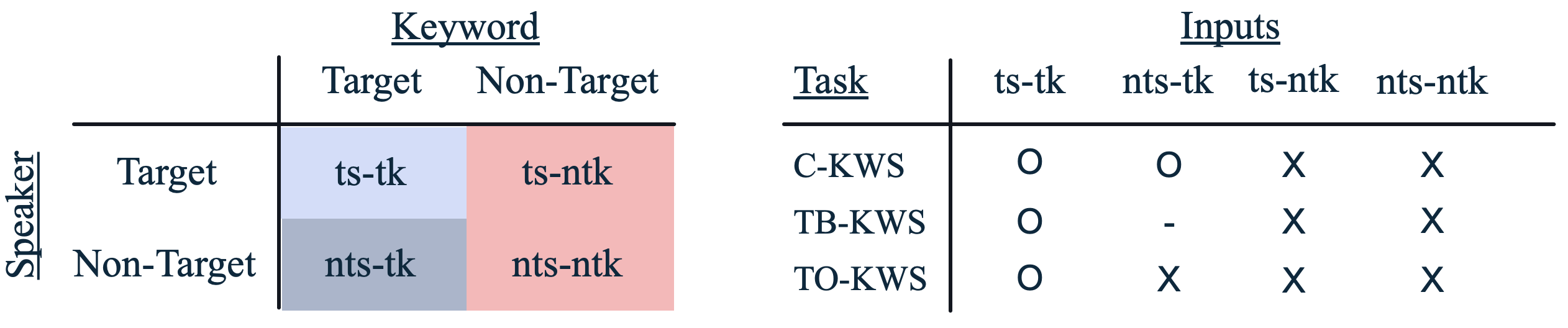,width=1.0\linewidth}
  \caption{}
  \label{figure1_task}
  \end{subfigure}
  
  \begin{subfigure}{1.0\linewidth}
  \centering
  \epsfig{figure=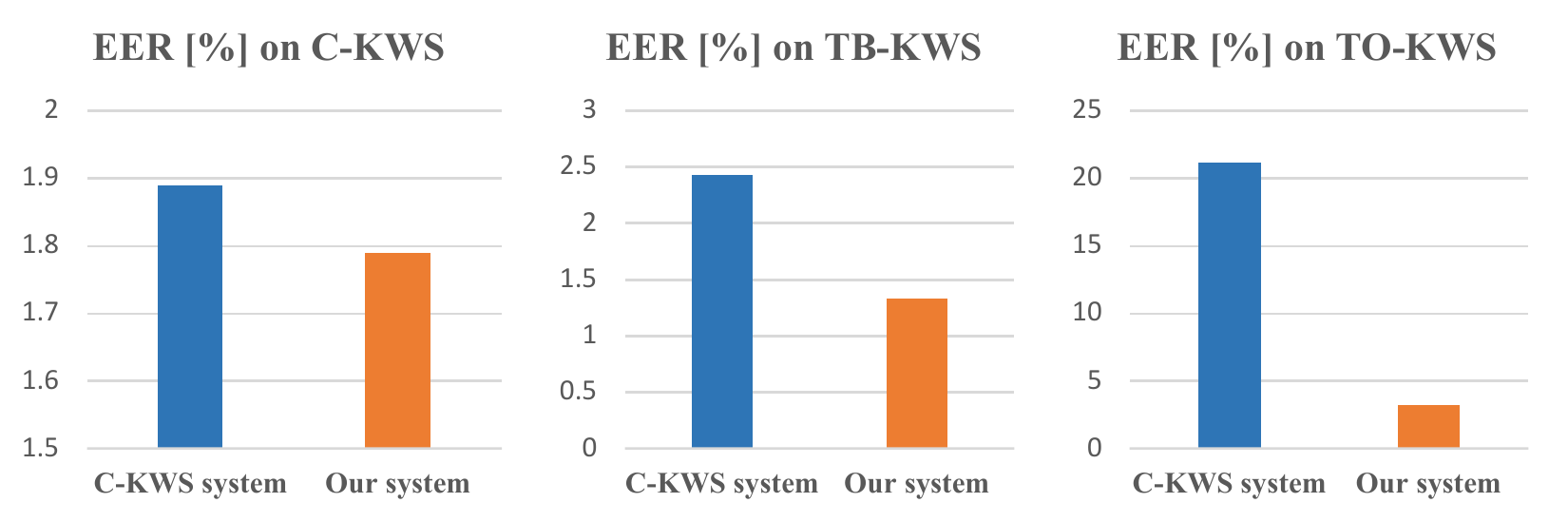,width=1.0\linewidth}
  \caption{}
  \label{figure1_performance}
  \end{subfigure}
  \vspace{-8pt}
\caption{(a): Inputs of keyword spotting systems can be categorized into four cases: ts-tk, nts-tk, ts-ntk, nts-ntk.
TB- and TO-KWS consider \textit{nts-tk} as neutral and negative, respectively.
(b): The conventional C-KWS system~\cite{bcresnet} shows low equal error rate (EER) on C-KWS, but it downgrades on two personalized tasks. Our system has a capability to conduct TB- and TO-KWS, while performing comparable result on C-KWS.}
\label{task:figure}
\vspace{-5pt}
\end{figure}

\begin{figure*}[!h]
\centering
  \epsfig{figure=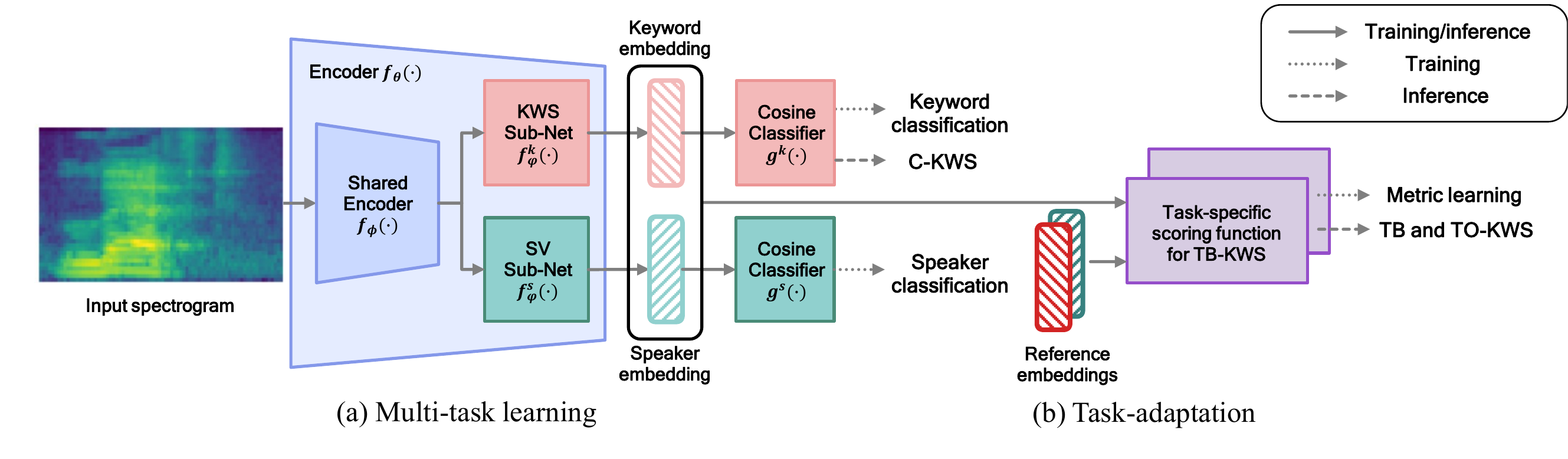,width=0.9\linewidth}
\vspace{-10pt}
\caption{\textbf{Personalized keyword spotting through multi-task learning} consists of two parts. (a) A multi-task learning part: the shared representation of keyword and speaker classification is encoded in low-level layers, and sub-networks learn the characteristics of each task in high-level features. (b) A task-adaptation part: initial representations are learned by the multi-task loss function, and these representations are adapted to each task through task-specific scoring functions.}
\label{fig:mainfigure}
\vspace{-10pt}
\end{figure*}

To describe personalized tasks, we categorize input utterances of the KWS system into four when a target speaker and a target keyword are given as Fig.~\ref{figure1_task} left. The four categories are \textit{ts-tk} (utterances from the target speaker and the target keyword), \textit{nts-tk} (utterances from a non-target speaker and the target keyword), \textit{ts-ntk}, and \textit{nts-ntk}.
The vague one is \textit{nts-tk}, which is the target keyword to detect but from a non-target user.
We design two personalized keyword spotting tasks considering \textit{nts-tk} differently as Fig.~\ref{figure1_task} right: Target user Biased KWS (TB-KWS) and Target user Only KWS (TO-KWS) to better focus on the target users. TB-KWS requires the model to be biased to the target users and does not explicitly consider \textit{nts-tk}. Moreover, some applications in devices prefer to be activated by only the target users, hence TO-KWS only detects \textit{ts-tk} and not \textit{nts-tk}. We evaluate the C-KWS system~\cite{bcresnet} on multiple KWS tasks in Fig.~\ref{figure1_performance}, and it shows performance degradation on our proposed personalized tasks since it does not consider user information.


We propose personalized keyword spotting through multi-task learning (PK-MTL) that can be selectively utilized on all three tasks, C-KWS, TB-KWS, and TO-KWS.
Our PK-MTL is a two\bg{-}stage system consisting of multi-task learning and task-adaptation.
First, we propose to apply multi-task learning on keyword spotting and speaker verification to leverage speaker information to the keyword spotting system. 
With a slight increase in the number of parameters, our keyword spotting system learns both keyword and speaker characteristics\bg{,} and we can get both representations in a computation\bg{-}efficient way.
Next, we introduce task-specific scoring functions to adapt learned representations to TB- and TO-KWS tasks.
Since the purpose of \bg{the} two tasks is different (see Fig.~\ref{figure1_task}), a function that combines the information from keyword and speaker representations suitable to each task is needed.
We propose two modules, Score Combination Module (SCM) and Task Representation Module (TRM), as task-specific scoring functions.
SCM is an optimization-free approach that \bg{directly} combines 
initial representations from the multi-task learning part.
The second approach is to construct a trainable Task Representation Module (TRM) that is designed to take the two representations and learn new task-specific representation\bg{s} for each task. 
With the aid of these modules, PK-MTL can be adapted to each task.

We evaluate on Google Speech Commands~\cite{google_commands}.
Upon the multiple keyword\bg{s} spotting backbones~\cite{res15,bcresnet,dsresnet}, our framework significantly boosts the performance of TB- and TO-KWS while performing comparable results of C-KWS, with a slight increase in the number of parameters and computation.
We also test PK-MTL in a practical setting, where negative samples continuously come from TV or other sources.
\is{Specifically}, we exploit WSJ-SI200~\cite{WSJ} and Librispeech~\cite{panayotov2015librispeech} as negative samples, and the results show that our system dramatically reduces the false alarm rate \sh{with the aid of task-specific scoring functions.}
\vspace{-2pt}
\section{Method}
As shown in Fig. \ref{fig:mainfigure}, the overall architecture of our system comprises two parts: multi-task learning on keyword spotting (KWS) and speaker verification (SV) and task-specific scoring functions to adapt to personalized KWS tasks.

\noindent \textbf{Notations.} \bg{The training data, $\mathcal{D}_\text{train}$, consists of labeled samples, $\{(x_i, y_i)\}_{i=1}^{|\mathcal{D}_\text{train}|}$, where $x_i$ is an input audio feature, and $y_i = (y^k_i, y^s_i)$. $y^k_i$ and $y^s_i$ are the corresponding keyword and speaker labels, respectively. To leverage speaker characteristics to the keyword spotting system, we design a multi-task learning architecture that comprises a shared encoder, $f_\phi(\cdot)$, sub-networks for keyword spotting and speaker verification, $f^k_{\varphi}(\cdot)$ and $f^s_{\varphi}(\cdot)$, and classifiers, $g^{k}(\cdot)$ and $g^{s}(\cdot)$.
}


\noindent \textbf{Decision process.}
\sh{Given a pre-defined target keyword, conventional keyword spotting system obtains a score $\psi_{i,\text{ref}}$ of a test sample $x_i$ belonging to the given target keyword using the trained keyword classifier~\cite{res15, tcresnet}.
The system accepts $x_{i}$ as the positive sample when $\psi_{i,\text{ref}}>\delta$, where $\delta$ is a threshold; otherwise $x_{i}$ is rejected as the negative sample.
In this paper, we additionally utilize an enrolled utterance of the target user $x_\text{ref}$ following personalized systems~\cite{Belli_2022_WACV, chung2020defence} and leverage the similarity of $x_{i}$ and $x_\text{ref}$ into the score $\psi_{i,\text{ref}}$.}

\vspace{-5pt}
\subsection{Multi-task learning for personalized keyword spotting}
\vspace{-2pt}
For a given backbone network, $f_\theta(\cdot)$, where $\theta = \{\phi, \varphi\}$, we adopt hard-parameter sharing~\cite{Caruana1993MultitaskLA, baxter1997bayesian} for low-level layers, {\it i.e.}, the shared encoder, $f_\phi(\cdot)$.
The shared encoder can learn complementary information from both KWS and SV, and the shared design is more efficient in terms of memory and computation than using separate task designs~\cite{kanakis2020reparameterizing, Sun_2021_ICCV}.
Since features for KWS and SV are adversarial in high-level concept\bg{s}~\cite{yun2019end}, {\it i.e.}, keyword features are speaker-agnostic and vice versa, sub-networks, $f^k_\varphi(\cdot)$ and $f^s_\varphi(\cdot)$, are added to learn the characteristics of each task.
Then, we get keyword and speaker features, $z_{i}^{k} = f^k_\varphi(f_\phi(x_{i}))$ and $z_{i}^{s} = f^s_\varphi(f_\phi(x_{i}))$.
On top of that, we apply cosine similarity based classifiers~\cite{liu2017sphereface, liu2020prototype} as follows:
\vspace{-1mm}
\begin{equation}
    g^{k}(z_{i}^{k}) = \softmax(w \cdot \cossim(z_{i}^{k}, W^{k}) + b),
    \vspace{-1mm}
\end{equation}
where $W^{k}$ is the learnable weight for keyword classification, $\cossim$ represents cosine similarity, $\cossim(a,b) = a\cdot b/(||a||\text{ }||b||)$, and $w$ and $b$ indicate scale and bias, respectively (for simplicity of notation, we omit the superscript $k$).
We define a keyword classification loss by minimizing the negative log probability of the true class:
\vspace{-1mm}
\begin{equation}
L_{k} = \sum_{i}{-y^k_i\log g^{k}(z^{k}_{i})}.
\vspace{-1mm}
\end{equation}
We get $L_{s}$ for SV in the same manner and combine two task-specific loss functions as below:
\vspace{-1mm}
\begin{equation}
    L_\text{mtl} = L_{k} + \lambda L_{s},
    \vspace{-1mm}
\label{eq:mtlloss}
\end{equation}
where $\lambda$ indicates the importance of speaker information.
Different from the goal of previous multi-task learning on KWS and SV~\cite{KWS_SV_MTL1, KWS_SV_MTL2} that focuses on boosting each task, our framework focuses more on leveraging user information for better\bg{-}personalized KWS.
After training the network with the loss in Eq.~\ref{eq:mtlloss}, we use learned representations for TB- and TO-KWS.

\subsection{Task-specific scoring functions for TB- and TO-KWS}
\label{score_function}

\noindent
\textbf{Score Combination Module (SCM).}
The first approach is to obtain two scores of keyword and speaker independently and combine them directly into a task-specific score for the decision process in each personalized task.
Given the enroll utterance $x_\text{ref}$ of a target user $y_{ref}^{s}$, keyword and speaker scores of the test input sample $x_{i}$ for the user $y_{t}^{s}$ are calculated by:
\begin{equation*}
    \psi_{i,\text{ref}}^{k} = \cossim(z_{i}^{k}, W^{k}_{\text{ref}}) \ \ \text{and} \ \ \psi_{i,\text{ref}}^{s} = \cossim(z_{i}^{s}, f^s_{\varphi}(f_\phi(x_\text{ref})),
\end{equation*}
where $\psi_{i,\text{ref}}^{k}$ is the cosine similarity score between the keyword embedding $z_{i}^{k}$ and the trained keyword classifier weight $W^{k}_{\text{ref}}$ that can be regarded as the most representative embedding of the pre-defined target keyword.
$\psi_{i,\text{ref}}^{s}$ is the score between the speaker embedding of the input $z_{i}^{s}$ and the embedding of the enrolled target user $f^{s}_\varphi(f_\phi(x_{\text{ref}}))$.
Note that target keywords are pre-defined, but speakers are not overlapped between training and test. Therefore, we use the reference embedding, \bg{$z_\text{ref}^s$}, of the target user at the test time.
\sh{We define $\text{SCM}(\cdot,\cdot;\alpha)$ that combines two scores, $\psi^{k}$ and $\psi^{s}$, into new score for each task, $\psi^\text{tb}$ and $\psi^\text{to}$.
$\text{SCM}$ can be any combination functions, but we use a simple linear combination function, $\alpha \cdot \psi^{k} + (1-\alpha) \cdot \psi^{s}$.}
The objective of keyword spotting is to minimize the false rejection rate (FRR) at the given false alarm rate (FAR).
Here, FAR is the percentage of negative samples being incorrectly accepted\bg{,} while FRR is that of positive samples being incorrectly rejected.
To achieve this goal, we obtain the parameters of SCM at the target FAR $c$ (\%) as follows:
\vspace{-1mm}
\begin{equation}
    \alpha^{*} = \underset{\alpha}{\argmin}  \text{FRR}(\text{SCM}(\psi^{k}, \psi^{s};\alpha)), \text{ s.t. FAR} = c.
\label{SCM}
\end{equation}
\vspace{-4mm}

\noindent
\textbf{Task Representation Module (TRM).} The second approach is to add a trainable neural network, $\text{TRM}_\text{tb}(\cdot, \cdot)$ and $\text{TRM}_\text{to}(\cdot, \cdot)$, whose inputs are keyword and speaker embeddings\bg{,} and an output is a task-specific embedding.
We apply metric learning loss with our proposed batch construction to train this module to form the discriminative embedding for TB- and TO-KWS.
We define positive and negative samples based on an anchor sample mimicking the test case of each task as described in Fig.~\ref{figure1_task}.
Then, we apply an angular prototypical loss function~\cite{chung2020defence} to make the positive samples closer and the negative samples apart from the anchor sample.
First, we measure the similarity between query samples and prototypes as follows:
\vspace{-1mm}
\begin{equation}
\label{task_specific_score}
    \psi_{i,j}^\text{tb} = \cossim(\text{TRM}_\text{tb}(z_{i}^{k}, z_{i}^{s}), \text{TRM}_\text{tb}(p_{j}^{k}, p_{j}^{s})),
\vspace{-1mm}
\end{equation}
where prototypes $p_{j}^{k}$ and $p_{j}^{s}$ are reference embeddings of the corresponding keyword and speaker of $j$-th sample, here we use learned classifiers' weights $W_{j}^{k}$ and $W_{j}^{s}$ as prototypes \sh{in training while the embedding of the enrolled speaker is used for $p_{j}^{s}$ in test.}
Then, we define a task-specific loss function as follows:
\vspace{-1mm}
\begin{equation}
    L_\text{tb} = -\frac{1}{N}\sum_{i=1}^{N}\frac{\exp(w \cdot {\psi^{tb}_{i,i}} + b)}{\sum_{j=1}^{N}\exp(w \cdot {\psi^{tb}_{i,j}} + b)}.
    \label{eq:metric-loss}
\vspace{-1mm}
\end{equation}
We apply the same task-specific loss function for TO-KWS, but negative samples are selected differently, {\it i.e.}, \textit{nts-tk} is selected for negative samples.
TRM can extract discriminative features for each task by minimizing Eq.~\ref{eq:metric-loss}.
We denote our mutli-task learning architecture with TRM as PK-MTL.

\noindent
\textbf{Inference time.} Given the test input $x_{i}$, we use the score from keyword embeddings, $\psi^{k}_{i,\text{ref}}$, for C-KWS and task-specific scores from SCM or TRM, $\psi^{tb}_{i,\text{ref}}$ and $\psi^{to}_{i,\text{ref}}$, for personalized tasks, as illustrated in Fig.~\ref{fig:mainfigure}.

\begin{table*}[]
\begin{center}
\caption{Comparison of Top-1 test accuracy (\%), EER (\%), and FRR (\%) at FAR $1\%$ and FAR $10\%$ of SV, C-KWS, TB-KWS, and TO-KWS on Google Speech Commands v1. Reported numbers are mean (std) over five trials.\label{main_table}}
\vspace{-8pt}
\resizebox{1.0\linewidth}{!}{
\begin{tabular}{lcc|cc|ccc|ccc|rr}
\toprule
 & & \multicolumn{1}{c}{SV} & \multicolumn{2}{c}{C-KWS} & \multicolumn{3}{c}{TB-KWS} & \multicolumn{3}{c}{TO-KWS} \\
\cmidrule{3-4}
\cmidrule{5-7} 
\cmidrule{8-11}
Method & Backbone & EER & Acc. & EER & FRR at 1\% &  FRR at 10\% & EER & FRR at 1\% & FRR at 10\% & EER & \#Mult & \#Param\\
\midrule
Vanilla & BC-ResNet-3~\cite{bcresnet} & - & 97.57 (0.04) & {\bf 1.92 (0.15)} & 2.85 (0.09) & 1.89 (0.20) & 2.47 (0.03) & 96.50 (0.07) & 64.32 (0.26) & 21.17 (0.08) & 16.7M  & 63.5k \\
Vanilla (+SV) & BC-ResNet-3~\cite{bcresnet} & {\bf 3.32 (0.10)} & 97.57 (0.04) & {\bf 1.92 (0.15)} & 2.85 (0.09) & 1.89 (0.20) & 2.47 (0.03) & 96.50 (0.07) & 64.32 (0.26) & 21.17 (0.08) & 22.5M & 82.4k \\
Naive MTL & BC-ResNet-3~\cite{bcresnet} & 3.36 (0.13) & {\bf 97.68 (0.18)} & 1.98 (0.07) & 2.82 (0.10) & 1.71 (0.32) & 2.44 (0.06) & 96.44 (0.09) & 64.26 (0.24) & 21.12 (0.09) & 17.5M & 80.2k \\
\midrule
PK-MTL w/ SCM-M  & BC-ResNet-3~\cite{bcresnet} & 3.36 (0.13) & {\bf 97.68 (0.18)} & 1.98 (0.07) & 2.99 (0.17) & 0.90 (0.09) & 2.12 (0.19) & 6.28 (0.27) & 2.73 (0.13) & 3.89 (0.06) & 17.5M & 80.2k  \\
PK-MTL w/ SCM-GS & BC-ResNet-3~\cite{bcresnet} & 3.36 (0.13) & {\bf 97.68 (0.18)} & 1.98 (0.07) & 2.75 (0.14) & 1.08 (0.52) & 2.40 (0.05) & {\bf 6.24 (0.42)} & 1.97 (0.70) & 3.76 (0.17) & 17.5M & 80.2k  \\
PK-MTL & BC-ResNet-3~\cite{bcresnet} & 3.36 (0.13) & {\bf 97.68 (0.18)} & 1.98 (0.07) & {\bf 2.58 (0.10)}  & {\bf 0.75 (0.25)} & {\bf 2.02 (0.22)} & 6.56 (0.20) & {\bf 1.52 (0.18)} & {\bf 3.37 (0.15)} & 17.5M & 82.0k \\ 
\midrule
\midrule
Vanilla & Res15~\cite{res15}& - & 95.82 (0.21) & 2.38 (0.10) & 4.17 (0.23) & 2.23 (0.17) & 3.03 (0.05) & 96.36 (0.28) & 63.83 (0.82) & 21.25 (0.08) & 966.3M & 241.1k \\
PK-MTL & Res15~\cite{res15}& 4.43 (0.13) & 96.30 (0.19) & 2.25 (0.13) & 3.30 (0.33) & 0.92 (0.08) & 2.14 (0.19) & 9.35 (0.46) & 2.56 (0.21) & 4.44 (0.18) & 1,040.6M & 262.3k \\ \midrule
Vanilla &DS-ResNet18~\cite{dsresnet} & - & 96.83 (0.21) & 1.85 (0.20) & 3.10 (0.17) & 1.33 (0.17) & 2.41 (0.17)  & 96.11 (0.13) & 63.76 (0.30) & 21.18 (0.07) & 305.6M & 79.9k \\
PK-MTL &DS-ResNet18~\cite{dsresnet} & 3.29 (0.07) & 96.85 (0.32) & 1.91 (0.21) & 2.46 (0.12) & 0.81 (0.15) & 1.61 (0.06) & 6.62 (0.37) & 1.90 (0.18) & 3.68 (0.24) & 326.1M & 89.9k \\ \midrule
Vanilla &BC-ResNet-8~\cite{bcresnet} & - & 98.01 (0.16) & 1.89 (0.17) & 2.69 (0.11) & 0.77 (0.41) & 2.43 (0.04) & 96.30 (0.32) & 63.20 (0.64) & 21.20 (0.09) & 91.9M & 386.9k \\
PK-MTL &BC-ResNet-8~\cite{bcresnet} & 2.58 (0.10) & 97.87 (0.12) & 1.79 (0.27) & 1.72 (0.10) & 0.47 (0.05) & 1.33 (0.04) & 5.01 (0.17) & 1.41 (0.28) & 3.22 (0.27) & 96.6M & 501.7k \\
\bottomrule
\end{tabular}
}
\end{center}
\vspace{-25pt}
\end{table*}

\begin{table}[]
\begin{center}
\caption{FAR (\%) at FRR $1\%$ and $5\%$ on WSJ and Librispeech datasets. Reported numbers are mean (std) over five trials.\label{sub_table2}}
\vspace{-8pt}
\resizebox{\linewidth}{!}{
\begin{tabular}{c|c|cc|cc}
\toprule
        &          & \multicolumn{2}{c}{GSC+WSJ} & \multicolumn{2}{c}{GSC+Librispeech} \\
Method  & Backbone & FAR at $1\%$ & FAR at $5\%$ & FAR at $1\%$ & FAR at $5\%$ \\ \midrule
Vanilla & Res15~\cite{res15} & 41.49 (2.37) & 1.31 (1.87) & 33.77 (2.34) & 0.51 (0.08) \\
Vanilla & DS-ResNet18~\cite{dsresnet} & 47.44 (5.76) & 0.25 (0.39) & 37.46 (9.05) & 0.48 (0.06) \\
Vanilla & BC-ResNet-3~\cite{bcresnet} & 46.61 (7.45) & 0.00 (0.01) & 27.59 (3.23) & 0.19 (0.02) \\
Vanilla & BC-ResNet-8~\cite{bcresnet} & 45.00 (4.98) & 0.00 (0.00) & 21.83 (2.79) & 0.12 (0.02) \\ \midrule
PK-MTL (TB) & BC-ResNet-3~\cite{bcresnet} & 5.58 (4.20) & 0.00 (0.00) & 2.69 (1.98) & 0.02 (0.01) \\
\textbf{PK-MTL (TO)} & BC-ResNet-3~\cite{bcresnet} & \textbf{0.15 (0.11) }& \textbf{0.00 (0.00)} & \textbf{0.15 (0.07)} & \textbf{0.00 (0.00)} \\ \midrule
PK-MTL (TB) & BC-ResNet-8~\cite{bcresnet} & 1.32 (1.55) & 0.00 (0.01) & 0.99 (0.36) & 0.01 (0.00) \\
\textbf{PK-MTL (TO)} & BC-ResNet-8~\cite{bcresnet} & \textbf{0.20 (0.30)} & \textbf{0.00 (0.01)} &\textbf{ 0.15 (0.06) }& \textbf{0.01 (0.00)} \\
\bottomrule
\end{tabular}
}
\end{center}
\vspace{-25pt}
\end{table}

\section{Experiments}
\subsection{Experimental Setup}
{\bf Datasets.}
We evaluate conventional and personalized KWS systems on Google Speech Commands v1~\cite{google_commands}.
The dataset contains 64,727 utterances of total 30 words from 1,881 speakers. We follow the conventional 12 class classification setting~\cite{google_commands}, which consists of ten classes of ``Yes," ``No," ``Up," ``Down," ``Left," ``Right," ``On," ``Off," ``Stop," and ``Go" with two additional classes ``Unknown" and ``Silence," which indicate remaining twenty words and no speech, respectively.
We divide the dataset into training, validation, and test in the same way as~\cite{tcresnet, res15, bcresnet, google_commands, rybakov2020streaming}.
For testing on TB- and TO-KWS, we make sample-to-sample pairs consisting of positive and negative pairs.
We randomly select an anchor sample and choose \textit{ts-tk}, \textit{nts-tk}, \textit{ts-ntk}, and \textit{nts-ntk} samples of the anchor.
To reduce the performance variation, we obtained 10 test splits where each test split contains 16,000 pairs.
We report the average performance from 10 test splits.
Note that ``Silence" and ``Unknown" classes can be selected for non-target keywords but not for target keywords.
Moreover, we make 10 test splits to validate speaker verification performance, which contain 160,000 speaker pairs \bg{that} are randomly sampled from the test set.

We also evaluate the models trained on Google Speech Commands in a practical scenario, where speeches continuously come from news or conversations containing words that are general negatives for the system.
We exploit the WSJ-SI200~\cite{WSJ} and Librispeech~\cite{panayotov2015librispeech} datasets as the negatives for the above scenario.
Negative samples from WSJ-SI200 are segmented from the whole audio stream into \bg{the} one-second-long following~\cite{kim2019query}.
For Librispeech, we segment the entire audio stream into one-second-long in the public noisy test set.
\sh{We make a test pair where one sample is from general negatives and the other is the target keyword with a randomly selected speaker from Google Speech Commands, and we conduct this process for all samples by pairing with $10$ target keywords.
}

\noindent {\bf Implementation details.}
We exploit three keyword spotting architectures as \is{the} backbone network, BC-ResNet~\cite{bcresnet}, Res15~\cite{res15} \is{and} DS-ResNet~\cite{dsresnet}, for our PK-MTL.
For BC-ResNet, we use input features of 40-dimensional log Mel spectrograms with 30 ms window length and 10 ms frameshift and apply data augmentations following~\cite{bcresnet}.
For \is{the} other two backbones, we follow~\cite{res15,dsresnet} to add noise and random shift to each segment. Then, we extract $40$ dimensional Melfrequency cepstrum coefficient features and use them as inputs.
We design the shared encoder using the backbones except the last two conv layers for BC-ResNet and the last conv block for Res15 and DS-ResNet.
Sub-networks follow the shared encoder, which consist of the remaining layers of each backbone and an additional fully connected layer.
We follow the official training strategy~\cite{res15, bcresnet, dsresnet} of each baseline network, {\it e.g.}, a learning rate and a mini-batch size.
We set $\lambda$ to 0.1 in Eq.~\ref{eq:mtlloss}.
For the baselines, {\it i.e.}, BC-ResNet, Res15, and DS-ResNet, we add an additional fully connected layer before the classifier and use the cosine classifier, $g^{k}(\cdot)$.
This modification improves the performance with additional parameters and computational cost.

We define Score Combination Module (SCM) as a linear combination function.
We adopt the attention modules~\cite{hu2018squeeze} for Task Representation Module (TRM).
TRM extracts the concatenated normalized keyword and speaker embeddings attended by the attention weights that are obtained by two fully connected layers whose intermediate feature size is $2$.
TRM is trained for 50 epochs using the same training strategy of the backbones.

\noindent {\bf Evaluation metric.}
\sh{We use false alarm rate (FAR), false rejection rate (FRR), and equal error rate (EER) at which FAR and FRR are the same.
Also, Top-1 test accuracy is used for evaluating multiple keywords classification (C-KWS).}

\vspace{-3pt}
\subsection{Ablation studies on Google Speech Commands}
\vspace{-2pt}
\noindent {\bf Effectiveness of multi-task learning.}
We compare three methods: 1) Vanilla is the conventional KWS model, 2) Vanilla (+SV) uses an additional single task model for SV, and 3) Naive MTL indicates our MTL framework trained by Eq.~\ref{eq:mtlloss}.
In Table~\ref{main_table}, the results of EER on SV and C-KWS demonstrate that the keyword spotting system can learn speaker representations without performance degradation on KWS with a marginal increase in the number of parameters and computation compared to Vanilla.
Vanilla (+SV) also leverages speaker information using single-task SV model, but it requires a higher computational cost due to separately performing each task.

\noindent {\bf Impact of task-specific scoring functions.}
To analyze the impact of scoring functions, we apply SCM and TRM to Naive MTL whose backbone is BC-ResNet-3.
With a combination function whose $\alpha$ is manually chosen as $0.5$, SCM-M, it cannot consistently improve TB-KWS even though speaker information is exploited to the task.
We can improve the performance at the target FAR, especially 1\%, by finding $\alpha$ through grid-search minimizing Eq.~\ref{SCM} at the target FAR on the validation set, and PK-MTL w/ SCM-GS largely reduces FRR at FAR 1\% of TB- and TO-KWS.
However, it has limited performance improvements because their representations cannot learn task-specific characteristics explicitly.
Therefore, we utilize TRM so that PK-MTL fully adapts keyword and speaker representations to TB- and TO-KWS, and it helps reducing FRR and EER significantly.

\noindent {\bf PK-MTL on various baselines.}
Our framework can be applied to any C-KWS architectures, hence we use three backbones, Res15, DS-ResNet18, BC-ResNet-8.
In Table~\ref{main_table}, even though the performance of SV and the influence of SV to KWS induced by multi-task learning are different depending on the structure and size of the backbones, PK-MTL improves the performance consistently on all personalized tasks with the aid of the speaker representations and the task-specific scoring function, TRM.

\begin{figure}[]
\vspace{-5pt}
\centering
  \epsfig{figure=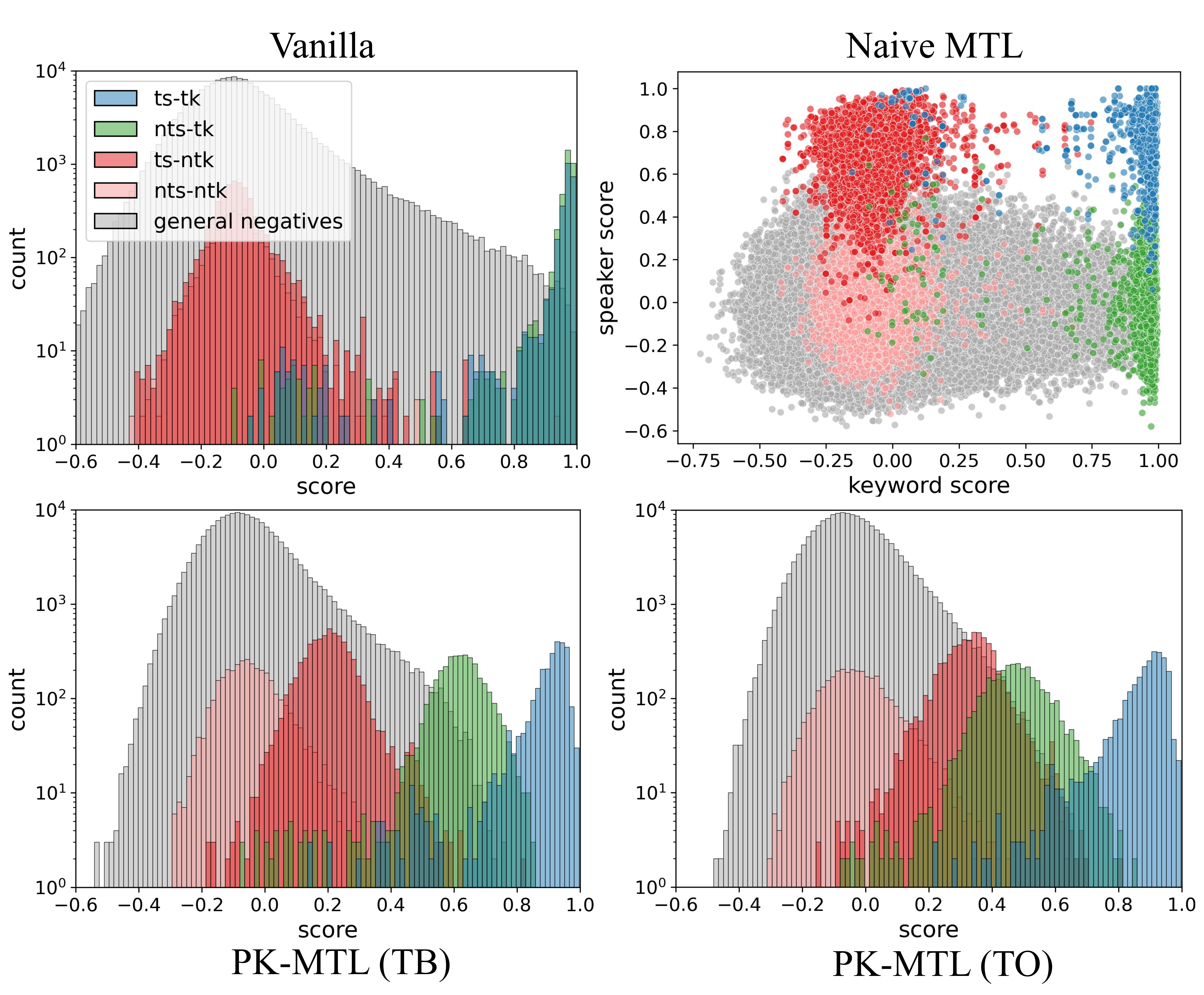,width=0.9\linewidth}
  \vspace{-7pt}
  \caption{The histograms (log scale) for Vanilla and PK-MTL show the distribution of scores from keyword and task-specific embeddings, respectively. The scatter plot for Naive MTL shows the distribution of keyword (x-axis) and speaker (y-axis) scores.}
  \label{figure_score}
\vspace{-20pt}
\end{figure}

\vspace{-4pt}
\subsection{Experiments on the realistic scenario}
\vspace{-2pt}
\sh{The KWS system trained on Google Speech Commands selects a threshold based on the target FRR of positive samples, {\it i.e.}, \textit{ts-tk}.
With the threshold, we evaluate KWS systems on WSJ and Librispeech datasets to validate whether they can reject general negatives.
}
All vanilla methods cannot reject general negatives that contain the target keywords, {\it e.g.}, ``Yes" and ``On," or similar keywords because speaker information is ignored for detecting keywords.
It makes KWS systems induce high FAR at the operating points of FRR $1\%$ as shown in Table~\ref{sub_table2}, and it can lead to increased power consumption.
However, in our systems, general negatives are easily rejected by considering speaker representations since the speaker characteristics of general negatives are different from the enrolled users.
Especially, PK-MTL (TO) can effectively reject general negatives even at the low operating point, FAR 1\%, because it learns to reject \textit{nts-tk}.

We analyze the score distribution of KWS systems as shown in Fig. 3.
We plot the scores of \textit{ts-tk}, \textit{nts-tk}, \textit{ts-ntk}, \textit{nts-ntk} pairs from Google Speech Commands and \sh{pairs of target keywords and general negatives from Librispeech.}
In Vanilla, the scores of target keywords and non-target keywords are well split regardless of speaker identities, but a lot of general negatives also result in high scores.
We plot both keyword and speaker scores through Naive MTL. As expected, general negatives have low speaker scores even though their keywords are similar or the same as the target.
The figure for PK-MTL (TB) shows that the score of \textit{ts-tk} is higher than those of \textit{nts-tk}, which indicates the model is biased toward the target users.
Moreover, \textit{ts-tk} is well separated from other samples in PK-MTL (TO), which objective is to accept \bg{the} target user only.
In PK-MTL, the scores of general negatives are apart from those of positive samples, thus we can reject them through the scores of TB or TO.

\vspace{-4pt}
\section{Conclusions}
\vspace{-1pt}
In this paper, we propose to leverage speaker information into the keyword spotting system to tackle personalized keyword spotting tasks.
Through proposed multi-task learning and task adaptation, our system can adapt to personalized tasks.
Extensive experiments show that leveraging speaker information reduces false alarm and rejection rate\bg{s} significantly in personalized and realistic keyword spotting scenarios.

\bibliographystyle{IEEEtran}

\bibliography{mybib}


\end{document}